\documentclass{article}
\usepackage{hyperref}  
\usepackage{doi}       
\usepackage{graphicx}
\usepackage{authblk}
\usepackage{amsmath}
\usepackage{amssymb}
\usepackage{float}
\usepackage[top=2cm, bottom=2cm, left=2.5cm, right=2cm]{geometry}
\usepackage[numbers,sort&compress]{natbib}
\usepackage{multirow}
\usepackage{longtable}
\usepackage{comment}
\usepackage{booktabs}
\usepackage[pagewise]{lineno}
\usepackage[cmyk]{xcolor}

\newcommand{\upcite}[1]{\textsuperscript{\cite{#1}}}
\title{Investigating Interacting Dark Energy Models Using Fast Radio Burst Observations}
\author[1]{Hang Yan}
\author[1]{Yu Pan\thanks{Corresponding author.Email:panyu@cqupt.edu.cn}}
\author[1]{Jia-Xin Wang}
\author[1]{Wen-Xiao Xu}
\author[1]{Ze-Hui Peng}
\affil[1]{School of Electronic Science and Engineering, Chongqing University of Posts and Telecommunications, Chongqing 400065, P. R. China}
\begin{document}
\maketitle
\begin{abstract}
This paper investigates the utility of Fast Radio Bursts (FRBs) as novel observational probes to constrain models of interacting dark energy (IDE). By leveraging FRBs' dispersion measures (DMs) and redshifts, we perform a comprehensive analysis of three IDE models—$\gamma_{m}$IDE, $\gamma_{x}$IDE, and $\xi$IDE—using Markov Chain Monte Carlo (MCMC) methods based on 86 localized FRBs and simulated datasets containing 2,500 to 10,000 mock events. By disentangling the contributions to the observed DMs from the Milky Way, host galaxies, and the intergalactic medium (IGM), key cosmological parameters are constrained, including the Hubble constant ($H_{0}$), matter density ($\Omega_{m}$), the dark energy equation of state ($\omega_{x}$), and interaction strengths ($\gamma_{m}$, $\gamma_{x}$, $\xi$). The best-fit values of the $\gamma_{m}$IDE models indicate a potential alleviation of the cosmic coincidence problem. Subsequently, we utilize information criteria (IC) to conduct a comparative assessment of the three IDE models. When applied to the current sample of observed FRBs, the $\xi$IDE model yields slightly lower IC values than the $\gamma_{m}$IDE and $\gamma_{x}$IDE models across all three information criteria, although the differences are not statistically significant. Notably, our study emphasizes the significance of current FRB observations in exploring potential interactions within the dark sector. These results underscore the value of FRB measurements as valuable complementary probes that provide further constraints on alternative cosmological models.

\end{abstract}
\noindent\textbf{Keywords:} Cosmological parameter estimation; Interacting dark energy models; Fast radio burst data
\section{Introduction}
Fast Radio Bursts (FRBs) are millisecond-duration radio transients characterized by unusually high dispersion measures (DMs), which serve as unique probes for studying the large-scale structure and evolution of the Universe\upcite{1}. Since their initial discovery by Lorimer et al.\upcite{2}, more than 800 FRB events have been detected, with some precisely localized to host galaxies at cosmological distances\upcite{3,2020Natur.581..391M}. The DM value of an FRB reflects the cumulative column density of free electrons along the line of sight, which can be decomposed into contributions from the Milky Way ($\mathrm{DM_{MW}}$), the host galaxy ($\mathrm{DM_{host}}$), and the intergalactic medium (IGM) contribution ($\mathrm{DM_{IGM}}$)\upcite{5}. By disentangling these components, FRB observations provide valuable constraints on $H_0$ and $\Omega_m$.
\par 
The cosmological utility of FRBs arises from the statistically significant correlation between the intergalactic medium's contribution to the dispersion measure ($\mathrm{DM_{IGM}}$) and redshift ($z$). Theoretical studies have shown that the mean value, $\langle \mathrm{DM_{IGM}} \rangle$, is directly related to the $H_0$ and the baryon density parameter ($\Omega_b$)\upcite{6}, In contrast, its distribution is notably affected by the inhomogeneity of baryon distribution in the intergalactic medium\upcite{7}.
\par
Recent studies have employed localized FRB data to measure the Hubble constant $H_0$\upcite{8}, and to test non-standard cosmological models, such as the $R_h=ct$ universe\upcite{9}. Moreover, several investigations have examined the impact of the FRB redshift distribution on cosmological parameter constraints\upcite{weihao}. In addition, FRB observations have been used to explore the possibility that compact objects (COs) constitute a fraction of dark matter\upcite{li}. FRBs also serve as effective probes for studying the expansion history of the Universe and the physical properties of the intergalactic medium\upcite{wh2,li2}. Notably, by breaking parameter degeneracies inherent in Cosmic Microwave Background (CMB) data, FRB observations can significantly tighten constraints on the Hubble constant and dark energy parameters\upcite{zx}. Furthermore, analysis of the CHIME/FRB catalog indicates that the observed redshift distribution of the FRB population cannot be fully explained by the star formation history (SFH) alone; instead, it requires the inclusion of a significantly delayed population or a mixture of populations to match the data\upcite{lin2024time,lin2024revised,zhang2022chime,qiang2022fast}. These findings provide a crucial theoretical foundation for our study and strongly motivate the use of FRBs as novel cosmological tools to constrain models, including those involving interactions between dark energy and dark matter.

\par
The physical nature of dark matter and dark energy, which together account for approximately 95\% of the Universe's energy density, remains one of the most fundamental challenges in modern cosmology. Although the standard Lambda Cold Dark Matter ($\Lambda$CDM) model successfully describes the accelerated expansion of the Universe, basic questions remain unanswered, such as the “coincidence problem” — the phenomenon that the current energy densities of dark matter and dark energy are of the same order of magnitude despite their very different evolutionary histories. To address this coincidence problem related to the evolution of dark energy and dark matter, researchers have proposed the IDE model, which incorporates a non-gravitational coupling term, $Q$, to characterize the energy exchange between dark matter and dark energy. As extensively discussed in the literature\upcite{10,11,12,13,14,15}, the IDE model assumes that the energy transfer between dark matter and dark energy can be described by an interaction term $Q$:

\begin{equation}\label{1}
    \dot{\rho}_{x} + 3H(\rho_{x} + p_{x}) = -Q \tag{1}
\end{equation}
\begin{equation}\label{2}
    \dot{\rho}_{m} + 3H\rho_{m} = Q \tag{2}
\end{equation}
where $\rho_x$ denotes the energy density of dark energy, $\rho_m$ denotes that of dark matter, and $p_x$ is the pressure of dark energy. The Hubble parameter $H$ characterizes the expansion rate of the Universe. The time derivatives $\dot{\rho}_x$ and $\dot{\rho}_m$ describe the evolution of the energy densities with cosmic time. The interaction term $Q$ quantifies the energy transfer between the two components: $Q > 0$ indicates a transfer of energy from dark energy to dark matter, whereas $Q < 0$ indicates the reverse. These equations ensure the conservation of the total energy density, as expressed by $\dot{\rho}_{x} + \dot{\rho}_{m} + 3H(\rho_{x} + \rho_{m} + p_{x}) = 0$.

This affects the integral form of $\langle \mathrm{DM_{IGM}} \rangle$. Meanwhile, the IDE model can also mitigate the Hubble tension to some extent\upcite{xu}. Current constraints on dark energy and dark matter interactions derived from Type Ia supernovae (SNe Ia), the CMB, and baryon acoustic oscillations (BAO) remain weak, primarily due to parameter parsimony and limited redshift coverage\upcite{18,19,20,21,pan2,p3}. Additionally, strong gravitational lensing time-delay measurements have emerged as a complementary tool to improve these constraints\upcite{pan}. In contrast, FRB data can effectively probe high-redshift ($z > 1$) regions, extending beyond the observational limitations of SNe Ia and providing new constraints on the redshift evolution of dark interactions. Moreover, unlike quasar absorption lines or X-ray surveys, FRBs offer a more comprehensive census of ionized baryons, including the diffuse IGM, which constitutes approximately 80\% of the total cosmic baryons\upcite{2020Natur.581..391M}.

\par
Previous studies have demonstrated the potential of FRB simulations in constraining parameters of IDE models\upcite{zhao2023probing}. This study employs real FRB data to investigate three specific IDE models: $\gamma_m$IDE, $\gamma_x$IDE, and $\xi$IDE. By imposing a relatively narrow Gaussian prior on \(\Omega_m\), the $\gamma_m$IDE model yields well-constrained posterior distributions for the remaining parameters. However, given the limitations of current observational data, the overall constraining power on cosmological parameters remains insufficient. To address this, we further analyze simulated FRB datasets and perform model selection using information criteria (IC), thereby establishing a more robust framework for future cosmological studies based on FRB observations.

\par
In \autoref{s2}, we provide a brief overview of DM measurements for localized FRBs. \autoref{s3} outlines the methodology and details the FRB samples employed in this study, including both observational and simulated data. In \autoref{s4}, we conduct Markov Chain Monte Carlo (MCMC) analyses to constrain three IDE models. \autoref{sec:Analysis} presents a theoretical interpretation and discussion of the results of the constraints. Finally, \autoref{s6} summarizes the main conclusions of the study.

\section{The Composition and Characterization of the Dispersion Volume Components in Fast Radio Bursts}\label{s2}
The precise localization of a group of FRBs within their host galaxies enables the establishment of a correlation between the dispersion measure (${\mathrm{DM_{FRB}}}$) and the redshift (${z_{\mathrm{FRB}}}$) of FRBs. ${\mathrm{DM_{FRB}}}$ represents the integral of the free electron density along the propagation path of an FRB signal and incorporates the effect of cosmic expansion on the path length by applying a redshift-dependent weighting factor of ${(1+z)^{\mathrm{{-1}}}}$.
\begin{equation}\label{3}
    \mathrm{DM}_{\mathrm{FRB}} = \int \frac{n_e \, dl}{1+z} \tag{3}
\end{equation}
From a physical perspective, the total ${\mathrm{DM_{FRB}}}$ can be decomposed into four primary components:

\begin{equation}\label{4}
    \mathrm{DM}_{\mathrm{FRB}}(z) = \mathrm{DM}_{\mathrm{MW}}^{\mathrm{ISM}} + \mathrm{DM}_{\mathrm{MW}}^{\mathrm{halo}} + \mathrm{DM}_{\mathrm{IGM}}(z) + \frac{\mathrm{DM}_{\mathrm{host}}}{1+z} \tag{4}
\end{equation}
The ${\mathrm{DM_{MW}^{ISM}}}$ component primarily originates from the interstellar medium (ISM) of the Milky Way and is typically estimated using Galactic electron density models such as the NE2001 model\upcite{24}. The ${\mathrm{DM_{MW}^{halo}}}$ term, which represents the contribution from the Milky Way's halo, remains poorly constrained but is generally expected to fall within the range of 50-80~pc~${\mathrm{cm^{-3}}}$. In the subsequent analyses of this paper, we adopt a fiducial value of ${\mathrm{DM_{MW}^{halo}}} = 65$~pc~${\mathrm{cm^{-3}}}$\upcite{prochaska2019probing}, noting that its variance and uncertainty are significantly smaller than those associated with ${\mathrm{DM_{IGM}}}$ and ${\mathrm{DM_{host}}}$, and can be effectively absorbed into the broader uncertainty budget of ${\mathrm{DM_{host}}}$\upcite{3, 25, 32}. 

\par
The ${\mathrm{DM_{IGM}}}$ component exhibits a pronounced dependence on redshift, as FRB signals traverse a considerable portion of the IGM during their propagation. The electron density in the IGM evolves with cosmic time, resulting in a substantial increase in ${\mathrm{DM_{IGM}}}$ with redshift. The redshift dependence of the mean ${\mathrm{DM_{IGM}}}$ can be quantitatively expressed by the following equation\upcite{26}:
\begin{equation}\label{5}
    \langle \mathrm{DM}_{\mathrm{IGM}}(z) \rangle = \frac{3cH_0\Omega_b f_d}{8\pi Gm_p} \int_0^z \frac{(1+z')\chi_e(z')}{E(z')} \, dz' \tag{5}
\end{equation}
where $c$ is the speed of light, $H_0$ is the Hubble constant, $\omega_b$ is the baryon density parameter, $G$ is the gravitational constant, $m_p$ is the proton mass, and $f_d \simeq 0.84$\upcite{27} represents the fraction of cosmic baryons residing in diffuse ionized gas. The number of free electrons per baryon is given by $\chi_e(z) = \frac{3}{4} \chi_{e,\mathrm{H}}(z) + \frac{1}{8} \chi_{e,\mathrm{He}}(z),$
where $\chi_{e,\mathrm{H}}(z)$ denotes the ionization fraction of hydrogen, and $\chi_{e,\mathrm{He}}(z)$ denotes that of helium. For redshifts $z \lesssim 3$, both hydrogen and helium are assumed to be fully ionized\upcite{28}, such that $\chi_{e,\mathrm{H}}(z) = 1$ and $\chi_{e,\mathrm{He}}(z) = 1$, yielding $\chi_e = \frac{7}{8}$. The function $E(z')$ denotes the dimensionless Hubble parameter at redshift $z'$, which is determined by the underlying cosmological model.

\par
The probability distribution function (PDF) of ${\mathrm{DM_{IGM}}}$ is derived from a theoretical treatment of the IGM and the Galactic halo\upcite{prochaska2019probing,29}.

\begin{equation}\label{6}
    P_{\mathrm{IGM}}(\Delta) = A\Delta^{-\beta} \exp\left[ -\frac{(\Delta^{-\alpha} - C_0)^2}{2\alpha^2\sigma_{\mathrm{DM}}^2} \right], \quad \Delta > 0 \tag{6}
\end{equation}
Here, $\Delta$ denotes the ratio of $\mathrm{DM_{IGM}}$ to its mean value $\langle \mathrm{DM_{IGM}} \rangle$, i.e., $\Delta = \mathrm{DM_{IGM}} / \langle \mathrm{DM_{IGM}} \rangle$. The parameter $\beta$ characterizes the internal density distribution of gas within the Galactic halo, while $\alpha$ governs the variance of the dispersion, thereby determining the shape of the probability distribution. When the variance is low, the distribution approximates a Gaussian; at high variance, it exhibits significant skewness. The best fit to the model is achieved with $\alpha = 3$ and $\beta = 3$\upcite{2020Natur.581..391M}. The normalization constant $A$ ensures that the total integral of the probability distribution equals 1, while $C_0$ ensures that the mean of the distribution satisfies $\langle \Delta \rangle = 1$. The parameter $\sigma_{\mathrm{DM}}$ represents the effective standard deviation, which quantifies the physical variance of the dispersion measure, influenced by both the gas distribution in the Galactic halo and intersections with large-scale cosmic structures.

\par
Using the IllustrisTNG simulations, Zhang et al.\upcite{30} derived the intrinsic distribution of $\mathrm{DM_{IGM}}$ across various redshifts. By fitting these distributions, we obtain the best-fit values of the parameters $A$, $C_0$, and $\sigma_{\mathrm{DM}}$ as functions of redshift, based on their well-established power-law dependencies on $z$. For computational efficiency, we interpolate these parameters at redshifts not explicitly sampled in the simulation, ensuring consistency with the original fitting results.
\par
The PDF of $\mathrm{DM_{host}}$ is primarily informed by cosmological models. In this study, we assume that $\mathrm{DM_{host}}$ follows a log-normal distribution\upcite{32}, which possesses two notable characteristics: (1) it is scale-invariant, and (2) it features a long tail extending towards higher values. This long-tailed nature allows for large $\mathrm{DM_{host}}$ values, potentially arising from regions with elevated electron densities within host galaxies or their surrounding environments.
\par
Specifically, we assume that the probability distribution of $\mathrm{DM_{host}}$ can be expressed as:
\begin{equation}\label{7}
    P_{\mathrm{host}}(\mathrm{DM}_{\mathrm{host}}) = \frac{1}{(2\pi)^{1/2} \mathrm{DM}_{\mathrm{host}} \sigma_{\mathrm{host}}} \exp\left[ -\frac{(\ln\mathrm{DM}_{\mathrm{host}} - \mu)^2}{2\sigma_{\mathrm{host}}^2} \right] \tag{7}
\end{equation}
Here, $\mu$ denotes the mean of $\ln(\mathrm{DM_{host}})$, and $\sigma_{\mathrm{host}}$ denotes its standard deviation. The median of the log-normal distribution is given by $e^{\mu}$, and the standard deviation is $\sqrt{e^{2\mu + \sigma^2_{\mathrm{host}}}(e^{\sigma^2_{\mathrm{host}}} - 1)}.$ Based on the IllustrisTNG simulations by Zhang et al.\upcite{22}, the $\mathrm{DM_{host}}$ contributions for both repeating and non-repeating FRBs under varying cosmic redshift conditions were successfully characterized. Localized FRBs can be classified into three categories according to their host galaxy properties: type I repeating FRBs (e.g., FRB 20121102A)\upcite{33}, type II repeating FRBs (e.g., FRB 20180916B)\upcite{34}, and type III non-repeating FRBs.

For each class, the median value of $\mathrm{DM_{host}}$, i.e., $e^{\mu_{\mathrm{host}}}$, increases with redshift and follows a power-law relation of the form: $e^{\mu_{\mathrm{host}}}(z) = \kappa (1 + z)^{\alpha},$ where the best-fit parameters $\kappa$ and $\alpha$ are provided in their study. In this work, we adopt the parameterization framework from Zhang et al.\upcite{22} to determine $\mu_{\mathrm{host}}$ and $\sigma_{\mathrm{host}}$ as functions of redshift, establishing a reliable basis for subsequent theoretical modeling. 

It should be noted that we adopt the redshift-dependent median $\mu_{\rm host}$ and dispersion $\sigma_{\rm host}$ of the log-normal $\mathrm{DM_{\rm host}}$ distribution from the IllustrisTNG simulation\upcite{22}. This approach may underestimate statistical uncertainties and introduce unknown systematic biases. A more reliable method would be to marginalize over these parameters or treat them as hyperparameters\upcite{2020Natur.581..391M}, which we leave for future work.

\section{Methods and Data}\label{s3}

The $\mathrm{DM}$-$z$ relation observed in these FRB samples can be utilized to test and compare various cosmological models. To constrain the cosmological parameters, we compute the joint likelihood across the full FRB dataset under each model. This approach allows for efficient optimization of model parameters by maximizing the likelihood function derived from the observed $\mathrm{DM}$–$z$ distributions.

\begin{equation}\label{8}
    \mathcal{L} = \prod_{i=1}^{N_{\mathrm{FRBs}}} P_i(\mathrm{DM}'_{\mathrm{FRB},i} | z_i) \tag{8}
\end{equation}
where $P_i(\mathrm{DM}'_{\mathrm{FRB},i} \mid z_i)$ denotes the probability of observing a corrected total dispersion measure, $\mathrm{DM}'_{\mathrm{FRB},i}$, for the $i$-th FRB after accounting for Galactic contributions.
\begin{equation}\label{9}
    \mathrm{DM}'_{\mathrm{FRB}} \equiv \mathrm{DM}_{\mathrm{FRB}} - \mathrm{DM}_{\mathrm{MW}}^{\mathrm{ISM}} - \mathrm{DM}_{\mathrm{MW}}^{\mathrm{halo}} = \frac{\mathrm{DM}_{\mathrm{host}}}{1+z} + \mathrm{DM}_{\mathrm{IGM}} \tag{9}
\end{equation}
\begin{equation}\label{10}
    \begin{aligned}
        P_i\left( \mathrm{DM}'_{\mathrm{FRB},i} | z_i \right) &= \int_0^{\mathrm{DM}'_{\mathrm{FRB},i}} P_{\mathrm{host}}(\mathrm{DM}_{\mathrm{host}}) \times \\
        &\quad P_{\mathrm{IGM}}(\mathrm{DM}'_{\mathrm{FRB},i} - \frac{\mathrm{DM}_{\mathrm{host}}}{1+z}, z_i) \, \mathrm{dDM}_{\mathrm{host}}
    \end{aligned}
    \tag{10}
\end{equation}
where $P_{\mathrm{host}}$ is the probability density function of $\mathrm{DM}{\mathrm{host}}$, and $P_{\mathrm{IGM}}$ is that of $\mathrm{DM}_{\mathrm{IGM}}$ (see \autoref{6} and \autoref{7}). We use the Python MCMC module EMCEE to explore the posterior probability distribution of the free parameters\upcite{37}. To determine the expected value $\langle \mathrm{DM}_{\mathrm{IGM}}(z) \rangle$ predicted by a given cosmological model, an expression for the dimensionless Hubble expansion rate $E(z)$ is required. The functional form of $E(z)$ will be discussed in the following section.
\par

This study employs both observational and simulated FRB samples for analysis. The sample of localized FRBs used in this study is summarized in \autoref{tab1}. We exclude FRB20220208A, FRB20220330D, FRB20221027A, and FRB20230216A from our analysis due to low confidence in their host galaxy associations, with host association probabilities \(P_{\mathrm{host}} < 90\%\)\upcite{sharma2024preferential}. Additionally, FRB20230718A, FRB20181220A, and several others are excluded because they do not satisfy the criterion \(\mathrm{DM}_{\mathrm{obs}} - \mathrm{DM}_{\mathrm{MW}}^{\mathrm{ISM}} > 80\, \mathrm{pc\, cm}^{-3}\), which ensures a significant extragalactic dispersion measure contribution.

\par Due to the limited number of FRBs with both reliable redshift measurements and well-constrained dispersion measures, as well as the uncertainties associated with their progenitors and host galaxy environments, current observational data remain insufficient to robustly constrain cosmological parameters. In particular, the existing sample lacks the statistical power to break degeneracies among key parameters such as the Hubble constant \(H_0\), the dark energy equation of state parameter \(\omega_x\), and the dark matter interaction parameter \(\gamma_m\)\upcite{35}. To overcome this limitation and assess the potential of FRBs as cosmological probes, we utilize simulated data to forecast their constraining power on cosmological models. This approach compensates for current observational gaps and provides a theoretical basis for optimizing future FRB surveys and their application in precision cosmology\upcite{36}.

\par
To more accurately model the redshift and dispersion measure distributions of FRBs, we adopt the following assumptions and procedures:

\begin{itemize}
    \item \textbf{Redshift distribution model:}  
    The intrinsic redshift distribution of FRBs is still uncertain due to the limited number of localized sources. Several models have been proposed, including distributions tracing the star formation rate or assuming a constant comoving density\upcite{li2019cosmology,zhang2021energy}. In this work, we adopt a GRB-like distribution \( P(z) \propto z e^{-z} \), as it naturally accounts for the observed decline in FRB detection rates at \( z > 1 \) and offers a simple, empirical form supported by earlier studies \upcite{9}.

    \item \textbf{Redshift range constraint:}  
    To ensure a well-defined ionization state and electron content, we restrict our analysis to the redshift range: $0 < z < 3$, where the free electron contribution from baryons can be reliably modeled.

    \item \textbf{Redshift sampling:}  
    The redshift values of the FRBs are drawn from the above PDF using the inverse transform sampling method.

    \item \textbf{Sampling of dispersion measures:}  
    The host galaxy and intergalactic medium contributions to the dispersion measure are generated as follows:
    \begin{itemize}
        \item Use \autoref{6}, which provides the PDF of $\mathrm{DM_{host}}$, and \autoref{7}, which provides that of $\mathrm{DM_{IGM}}$.
        
        \item To compute $\mathrm{DM_{IGM}}$, a cosmological baseline model is required. We adopt the standard flat $\Lambda$CDM model, with cosmological parameters set according to the latest Planck results, specifically $H_0 = 67.4\ \mathrm{km\ s^{-1}\ Mpc^{-1}}$ and $\Omega_m = 0.317$\upcite{18}. In each of the three IDE models introduced in Section~\ref{s4}, the model reduces to the standard $\Lambda$CDM scenario under a specific condition on its interaction parameter and the dark energy equation-of-state: $\gamma_m = 0$ and $\omega_x = -1$ for the $\gamma_m$IDE model, $\gamma_x = 0$ and $\omega_x = -1$ for the $\gamma_x$IDE model, and $\xi = 3$ and $\omega_x = -1$ for the $\xi$IDE model. These fiducial values are adopted as baseline inputs in our mock data generation.
        
        \item The distribution of $\mathrm{DM_{host}}$ is modeled using a log-normal distribution characterized by parameters $\mu$ and $\sigma$, where the median value is given by $e^{\mu}$. Different combinations of $\mu$ and $\sigma$ are used to simulate a range of possible $\mathrm{DM_{host}}$ scenarios, reflecting astrophysical uncertainties in host galaxy electron density environments. For simplicity, we adopt a uniform model for all simulated FRBs with $e^{\mu} = 32.97 \times (1 + z)^{0.84}$\upcite{22}.
        
        \item Apply inverse transform sampling to generate $\mathrm{DM_{host}}$ and $\mathrm{DM_{IGM}}$ values for each sampled redshift, based on their respective PDFs.
    \end{itemize}

    \item \textbf{Total dispersion measure synthesis:}  
    With the mock $\mathrm{DM_{host}}$ and $\mathrm{DM_{IGM}}$, we calculate $\mathrm{DM}'_{\mathrm{FRB}}$ according to \autoref{9}.
\end{itemize}
These modeling assumptions enable the generation of realistic mock FRB samples, which are subsequently used to evaluate the potential of FRBs in constraining cosmological parameters. To investigate the performance of our model selection approach, we conducted two sets of simulations. First, we produced a sample of 2,500 FRBs.The goal was to determine cases where model discrimination can be achieved via the IC. Specifically, with a sample size of 2,500, the three IDE models can be strictly distinguished from the $\Lambda$CDM model based on the IC judgments. Then, to examine the effects of increasing the sample size by an order of magnitude, we simulated an additional 10,000 FRBs. This allows us to assess how model discrimination improves with larger datasets and to compare trends observed in the smaller sample.

\section{Models and Constraining Results}\label{s4}
We find that \autoref{1} and \autoref{2} take the form of the continuity equation. Therefore, the interaction term describing the coupling between dark energy and dark matter must be proportional to the energy density and a quantity with dimensions of inverse time. In this study, we use the Hubble parameter $H$ to construct this relationship. Several standard formulations for the interaction term can be obtained by varying the form of the energy density \upcite{38,39}. The first interaction term is defined as:

\begin{equation}\label{11}
    Q_1 = 3\gamma_m H\rho_m \tag{11}
\end{equation}
In this expression, $\gamma_m$ is a constant that quantifies the interaction strength between dark matter and dark energy, and $\rho_m$ represents the dark matter energy density. This form implies that the interaction strength is proportional to the dark matter energy density and the Hubble parameter. The second interaction term is given by:

\begin{equation}\label{12}
    Q_2 = 3\gamma_x H\rho_x \tag{12}
\end{equation}
Here, $\gamma_x$ is a constant quantifying the interaction strength, and $\rho_x$ is the energy density of dark energy. This indicates that the interaction is proportional to the dark energy density and the Hubble parameter.

\par
Another interaction term is based on the scaling relation $\frac{\rho_x}{\rho_m} = \frac{\rho_{x0}}{\rho_{m0}} a^\xi$ \upcite{40}, which describes the ratio of dark energy density to matter density. Additionally, the third interaction term was proposed by Dalal et al.\upcite{40} and Guo et al.\upcite{41} within the framework of the flat Friedman-Lemaître-Robertson-Walker (FRW) model of the Universe:

\begin{equation}\label{13}
    Q_3 = \frac{-(1-\Omega_m)(\xi + 3\omega_x)}{1 - \Omega_m + \Omega_m(1+z)^\xi} H\rho_m \tag{13}
\end{equation}
Here, $\mathrm{\Omega_m}$ is the current matter density parameter, and $\mathrm{\omega_x} \equiv p/\rho$ is the equation of state (EoS) parameter of dark energy, representing the ratio of dark energy pressure to its energy density. In the spatially flat FRW model, $\mathrm{\omega_x}$ is assumed to be constant.

\par
Among these three interaction models, when $\gamma_m = 0$, $\gamma_x = 0$, and $\xi + 3 \omega_x = 0$, the model corresponds to the standard cosmological scenario with no interaction between dark energy and dark matter. Conversely, when $\gamma_m \neq 0$, $\gamma_x \neq 0$, and $\xi + 3 \omega_x \neq 0$, it represents a non-standard cosmological scenario with interactions. The $\Lambda$CDM limit is recovered in all three models when $\omega_x = -1$, specifically for $\gamma_m = 0$ in the first model, $\gamma_x = 0$ in the second, and $\xi = 3$ in the third. Moreover, $\gamma_m < 0$ or $\gamma_x < 0$ (with $\xi + 3 \omega_x > 0$) corresponds to energy transfer from dark matter to dark energy, while $\gamma_m > 0$ or $\gamma_x > 0$ (with $\xi + 3 \omega_x < 0$) suggests energy transfer from dark energy to dark matter. The magnitudes of $\gamma_m$ and $\gamma_x$ govern both the strength and direction of the interaction, while $\xi$ quantifies the severity of the coincidence problem.

Having introduced the adopted models, methodologies, and their associated parameters, we now proceed to specify the prior ranges for the free parameters employed in our model. Table~\ref{tab:priors} summarizes these parameters along with their adopted priors. In particular, the prior on $\Omega_m$ is modeled as a Gaussian distribution with mean $\mu = 0.317$ and standard deviation $\sigma = 0.007$, consistent with the latest Planck results\upcite{18}. This choice reflects the strong constraints on $\Omega_m$ provided by current cosmological observations, thereby effectively incorporating prior knowledge into our analysis.
\begin{table}[htbp]
    \centering
    \caption{The prior distributions of cosmological parameters}
    \label{tab:priors}
    \begin{tabular}{cc}
    \toprule[1.0pt]
   Parameter & Prior of parameter inference \\
    \hline
     $H_0$ & Uniform\([0, 100]\) \\
     $\Omega_m$ & Gaussian\([0.317,0.007]\) \\
     $\omega_x$ & Uniform\([-2, 0]\) \\
     $\gamma_m$ & Uniform\([-2, 2]\) \\
     $\gamma_x$ & Uniform\([-2, 2]\) \\
     $\xi$ & Uniform\([0, 6]\) \\
    \bottomrule[1.0pt]
    \end{tabular}
\end{table}

\subsection{The $\gamma_m$IDE model}
When the interaction between dark energy and dark matter is given by $Q_1 = 3\gamma_m H \rho_m$, we obtain the dimensionless Hubble parameter ($H(z)/H_0$):

\begin{equation}\label{14}
    E^2(z) = \frac{\omega_x \Omega_m}{\gamma_m + \omega_x} (1+z)^{3(1-\gamma_m)} + \left( 1 - \frac{\omega_x \Omega_m}{\gamma_m + \omega_x} \right)(1+z)^{3(1+\omega_x)} \tag{14}
\end{equation}
\par
\begin{figure}[H]
    \centering
    \begin{minipage}[t]{0.45\textwidth}
        \centering
        \includegraphics[width=\linewidth]{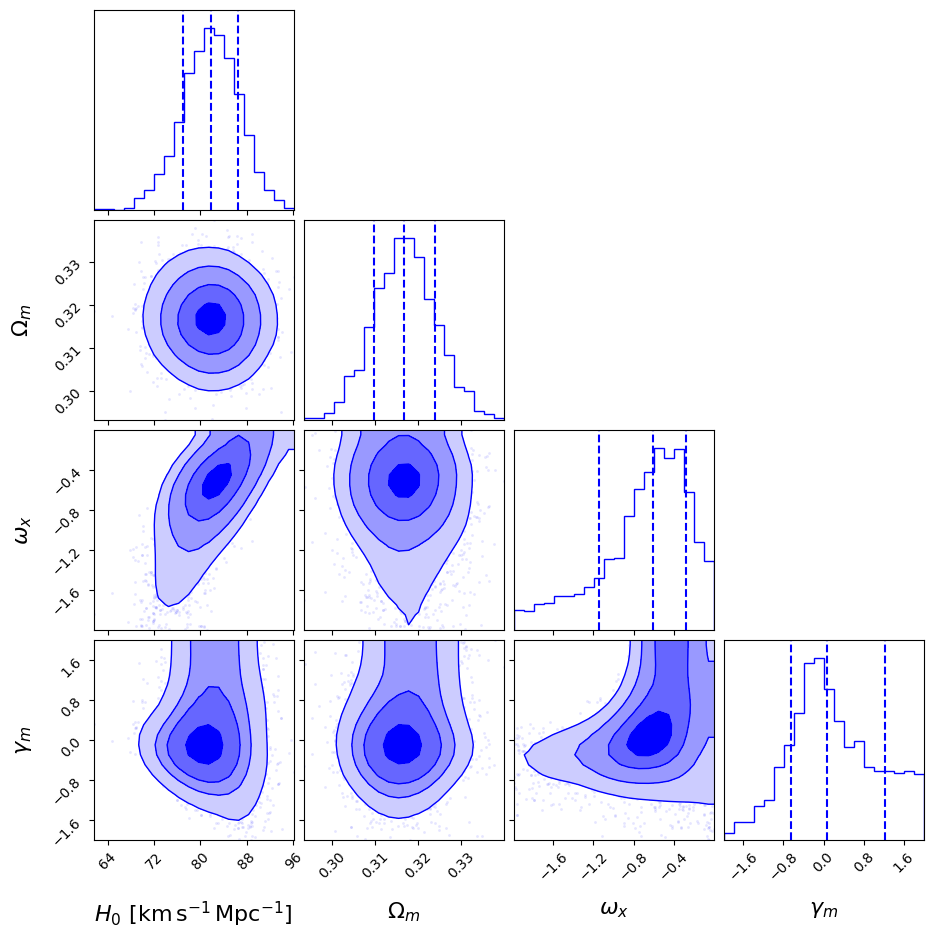}
    \end{minipage}
    
    \vspace{0.8em}  
    
    \begin{minipage}[t]{0.45\textwidth}
        \centering
        \includegraphics[width=\linewidth]{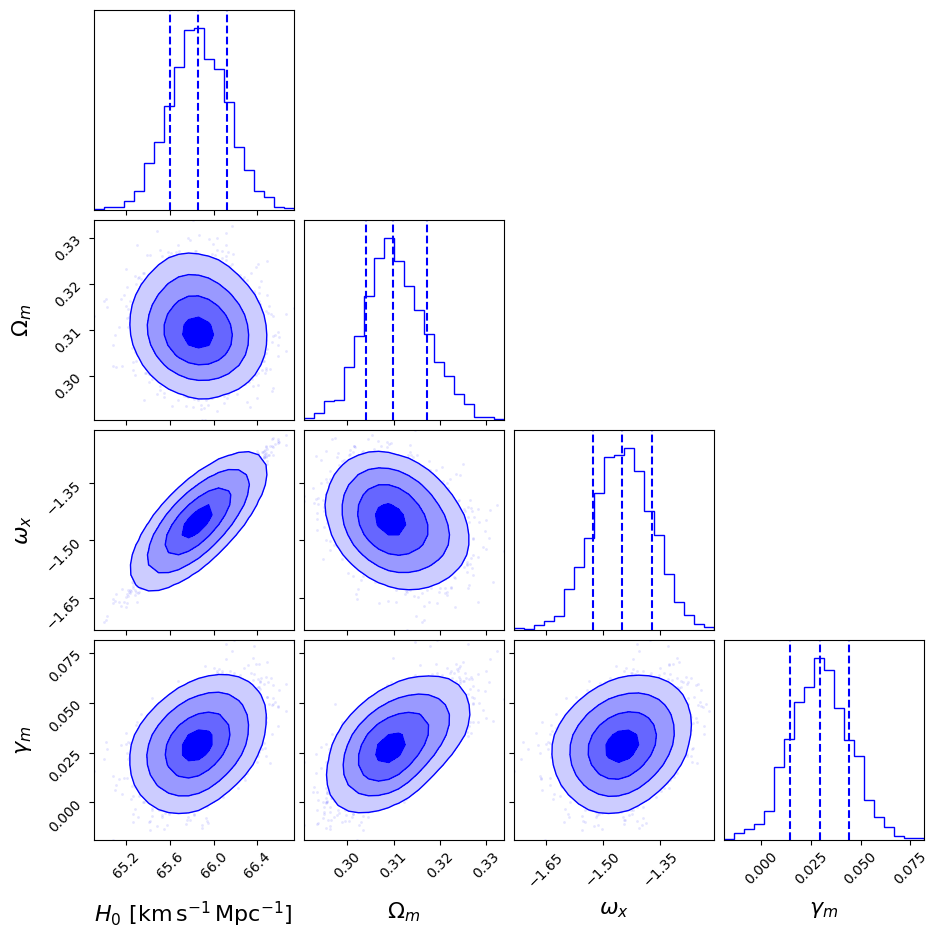}
    \end{minipage}
    \hfill
    \begin{minipage}[t]{0.45\textwidth}
        \centering
        \includegraphics[width=\linewidth]{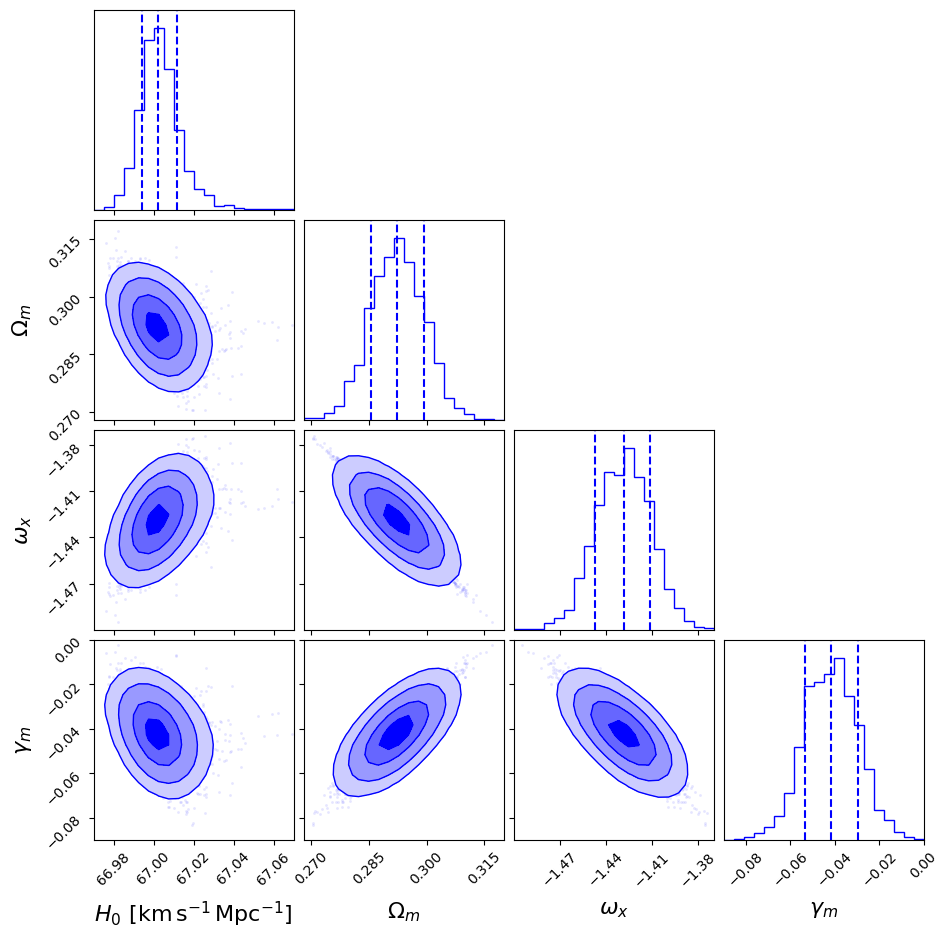}
    \end{minipage}
    
    \caption{Triangle plots showing constraints under the $\gamma_m$IDE model. Top: constraints from real FRB observations. Bottom left: constraints from 2500 simulated FRBs. Bottom right: constraints from 10000 simulated FRBs. In each subfigure, the dashed lines from left to right correspond to the 16\%, 50\%, and 84\% quantiles of the distribution.} The diagonal shows marginalized posterior distributions for each parameter, while off-diagonal panels show joint distributions between parameter pairs.
    \label{f1}
\end{figure}

 Here, $\Omega_m = 8 \pi G \rho_{m0}/{(3 H_0^2)}$ is the current fractional energy density of matter. In our analysis, a Gaussian prior of $\Omega_m = 0.317 \pm 0.007$ is adopted, consistent with recent cosmological observations, to constrain its contribution. The remaining free parameters are $H_0$, $\omega_x$, and $\gamma_m$. Given the defined likelihood function, we employ the MCMC method to explore the posterior probability distributions of these three free parameters. We employ a screened set of 86 observational FRBs, together with two simulated datasets comprising 2,500 and 10,000 FRBs. The resulting posterior distributions are shown in \autoref{f1}. When 86 observational FRB data constraints are applied, the best-fit values are determined as follows: the Hubble constant $H_0 = 81.81^{+4.62}_{-4.88}\ \mathrm{km\ s^{-1}\ Mpc^{-1}}$, the dark energy equation of state parameter $\omega_x = -0.61^{+0.33}_{-0.53}$, and the interaction parameter $\gamma_m = 0.64^{+1.15}_{-0.71}$ at 68.3\% confidence level.

\par 
Although the best-fit value of the interaction parameter $\gamma_m$ between dark energy and dark matter is $\gamma_m = 0.64^{+1.15}_{-0.71}$, indicating a mild preference for energy transfer from dark energy to dark matter and potential mitigation of the coincidence problem, the null-interaction case ($\gamma_m = 0$) still lies well within the 1$\sigma$ confidence interval. This suggests that current data remain consistent with both interacting and non-interacting dark sector scenarios. The best-fit value of the dark energy equation-of-state parameter is slightly greater than $-1$, indicating a mild preference for quintessence-like behavior. Both the estimation and uncertainty of $\omega_x$ in the $\gamma_m$IDE model are consistent with the $\Lambda$CDM framework ($\omega_x = -1$), suggesting that the current data are not yet sufficient to distinguish the $\gamma_m$IDE model from $\Lambda$CDM.

For subsequent model comparison in Section~\ref{sec:Analysis}, we record the best-fit log-likelihood $\ln \mathcal{L}_{\mathrm{max}}$ and the number of free parameters $k=4$ for BIC evaluation.

\begin{table}[H]
    \begin{center}\caption{Best-fit values and 1$\sigma$ uncertainties of the cosmological parameters in the three IDE models ($\gamma_m$IDE, $\gamma_x$IDE, and $\xi$IDE) from FRB datasets: 86 real observations and two simulated samples (2,500 and 10,000 mock FRBs).}
        \label{1t}
        \begin{tabular}{cccc} 
            \toprule[1.0pt]
            \multicolumn{4}{c}{\rule{0pt}{15pt}The $\gamma_m$IDE Model}     \\ [1pt]
            \cline{2-4}
             &  \rule{0pt}{10pt}$H_0$ $(\mathrm{km\,s^{-1}\,Mpc^{-1}})$  & $\omega_x$ & $\gamma_m$  \\ [1pt]
            \hline
            Observational FRB data & \rule{0pt}{15pt}$81.81^{+4.62}_{-4.88}(1\sigma)$  & $-0.61^{+0.33}_{-0.53}(1\sigma)$ & $0.64^{+1.15}_{-0.71}(1\sigma)$  \\ [5pt]
            Simulated FRB data(2500)& \rule{0pt}{15pt}$65.856^{+0.258}_{-0.255}(1\sigma)$  & $-1.449^{+0.075}_{-0.073}(1\sigma)$ & $0.0293^{+0.014}_{-0.015}(1\sigma)$  \\ [5pt]
            Simulated FRB data(10000)& \rule{0pt}{15pt}$67.001^{+0.010}_{-0.008}(1\sigma)$  & $-1.428^{+0.018}_{-0.018}(1\sigma)$ & $-0.0416^{+0.012}_{-0.012}(1\sigma)$  \\ [5pt]
            \hline
            \multicolumn{4}{c}{\rule{0pt}{15pt}The $\gamma_x$IDE Model}     \\ [1pt]
            \cline{2-4}
            &\rule{0pt}{10pt}$H_0$ $(\mathrm{km\,s^{-1}\,Mpc^{-1}})$  & $\omega_x$ & $\gamma_x$  \\ [1pt]
            \hline
            Observational FRB data & \rule{0pt}{15pt}$81.03^{+5.24}_{-4.88}(1\sigma)$  & $-1.03^{+0.65}_{-0.63}(1\sigma)$ & $-0.80^{+0.75}_{-0.80}(1\sigma)$  \\ [5pt]
            Simulated FRB data(2500)& \rule{0pt}{15pt}$66.055^{+0.247}_{-0.251}(1\sigma)$  & $-1.403^{+0.071}_{-0.075}(1\sigma)$ & $0.062^{+0.018}_{-0.018}(1\sigma)$  \\ [5pt]
            Simulated FRB data(10000)& \rule{0pt}{15pt}$66.982^{+0.013}_{-0.008}(1\sigma)$  & $-1.495^{+0.016}_{-0.015}(1\sigma)$ & $0.002^{+0.015}_{-0.017}(1\sigma)$  \\ [5pt]
            \hline
            \multicolumn{4}{c}{\rule{0pt}{15pt}The $\xi$IDE Model}     \\ [1pt]
            \cline{2-4}
            &\rule{0pt}{10pt}$H_0$ $(\mathrm{km\,s^{-1}\,Mpc^{-1}})$  & $\omega_x$ & $\xi$  \\ [1pt]
            \hline
            Observational FRB data & \rule{0pt}{15pt}$81.14^{+5.11}_{-5.60}(1\sigma)$  & $-0.72^{+0.41}_{-0.48}(1\sigma)$ & $3.44^{+1.77}_{-2.19}(1\sigma)$  \\[5pt]
            Simulated FRB data(2500)& \rule{0pt}{15pt}$65.978^{+0.248}_{-0.287}(1\sigma)$  & $-1.424^{+0.072}_{-0.086}(1\sigma)$ & $3.937^{+0.304}_{-0.271}(1\sigma)$  \\ [5pt]
            Simulated FRB data(10000)& \rule{0pt}{15pt}$67.701^{+0.012}_{-0.006}(1\sigma)$  & $-1.008^{+0.013}_{-0.012}(1\sigma)$ & $3.039^{+0.051}_{-0.047}(1\sigma)$  \\ [5pt]
            \bottomrule[1.0pt]
            \end{tabular}
    \end{center}
\end{table}
While current observational data provide useful constraints on the cosmological parameters, the relatively large uncertainties highlight the need for larger and more precise datasets. To further assess the constraining power of future FRB observations, we extend our analysis by incorporating two simulated FRB samples with significantly larger sizes. These mock datasets, consisting of 2,500 and 10,000 FRBs, are designed to mimic the statistical properties of real FRBs while significantly improving the data volume. Constraints based on the current sample of 86 localized FRBs provide a Hubble constant of $H_0 = 81.81^{+4.62}_{-4.88}~\mathrm{km\,s^{-1}\,Mpc^{-1}},$ corresponding to an uncertainty of approximately \(\pm 4.75~\mathrm{km\,s^{-1}\,Mpc^{-1}}\).
The intermediate simulated sample of 2,500 FRBs reduces this uncertainty to $\pm 0.26~\mathrm{km\,s^{-1}\,Mpc^{-1}},$
while the larger sample of 10,000 FRBs further reduces the uncertainty to $\pm 0.01~\mathrm{km\,s^{-1}\,Mpc^{-1}}.$ These results demonstrate the essential role of increased data volume in suppressing statistical fluctuations.

Similarly, the dark energy equation of state parameter \(\omega_x\), currently constrained to \(\omega_x = -0.61^{+0.33}_{-0.53}\) with an absolute uncertainty of approximately \(\pm 0.43\), sees its uncertainty reduced to \(\pm 0.07\) for the intermediate sample and further compressed to \(\pm 0.02\) for the large sample. As the sample size increases, the results exhibit clear convergence, indicating improved stability and reliability in the estimation of cosmological parameters.

For the coupling parameter $\gamma_m$, the current broad constraint of $0.64^{+1.15}_{-0.71}$ ($\pm 0.93$) narrows significantly to $\pm 0.015$ with the intermediate sample, and further tightens to $\pm 0.012$ for the large sample, indicating improved parameter stability. The slight deviation of $\omega_x$ from the standard cosmological value in the posterior results may be attributed to residual parameter degeneracies, which can shift the best-fit values even when statistical uncertainties are significantly reduced.

\subsection{The $\gamma_x$IDE Model}
By performing a similar analysis as before using the interaction term $Q_2 = 3\gamma_x H\rho_x$, which is proportional to the dark energy density, we obtain the Hubble parameter:

\begin{equation}\label{15}
    E^2(z) = (1 - \Omega_m)(1+z)^{3(1+\gamma_x+\omega_x)} + \frac{\omega_x \Omega_m + \gamma_x + \gamma_x(\Omega_m - 1)(1+z)^{3(\gamma_x + \omega_x)}}{(1+z)^{-3}(\gamma_x + \omega_x)} \tag{15}
\end{equation}

\begin{figure}[H]
    \centering
    \begin{minipage}[t]{0.45\textwidth}
        \centering
        \includegraphics[width=\linewidth]{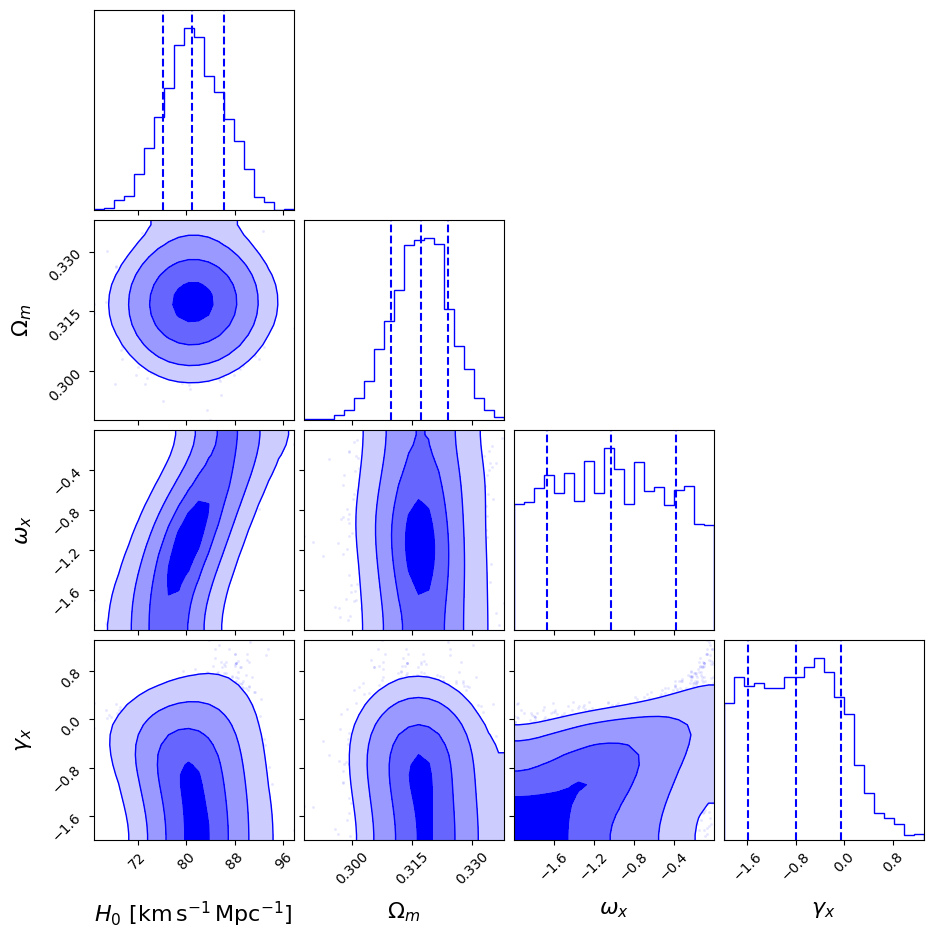}
    \end{minipage}
    
    \vspace{0.8em}  
    
    \begin{minipage}[t]{0.45\textwidth}
        \centering
        \includegraphics[width=\linewidth]{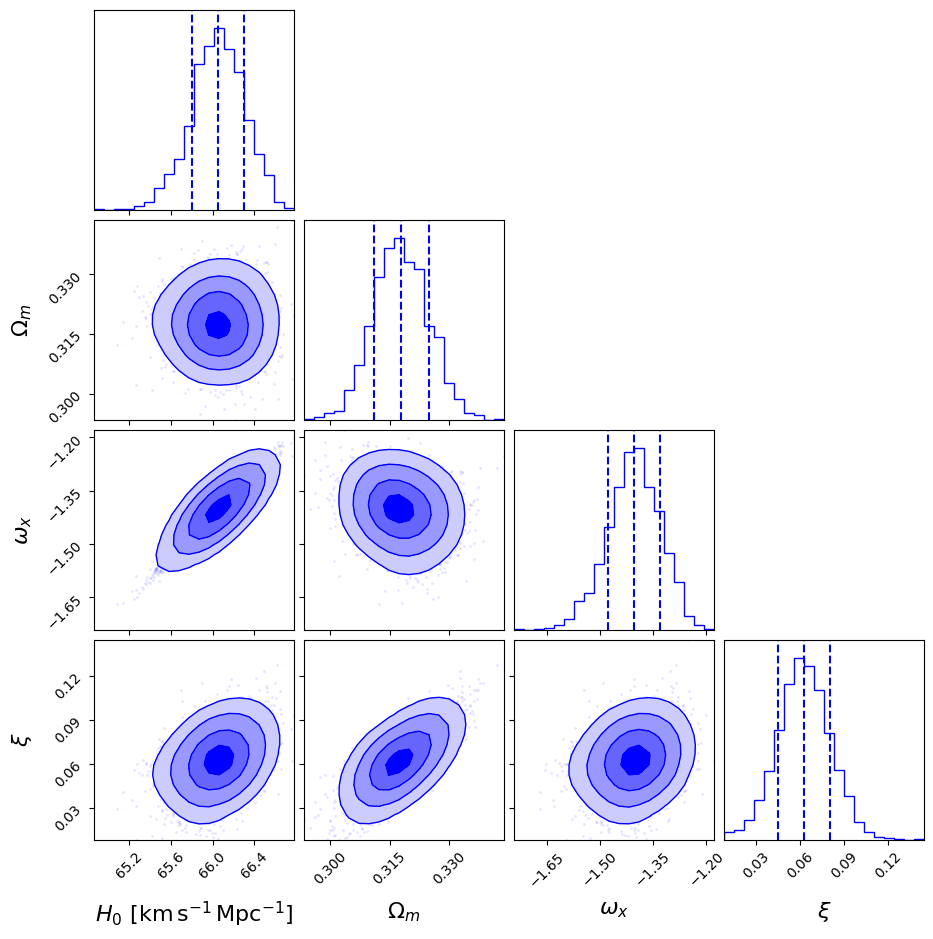}
    \end{minipage}
    \hfill
    \begin{minipage}[t]{0.45\textwidth}
        \centering
        \includegraphics[width=\linewidth]{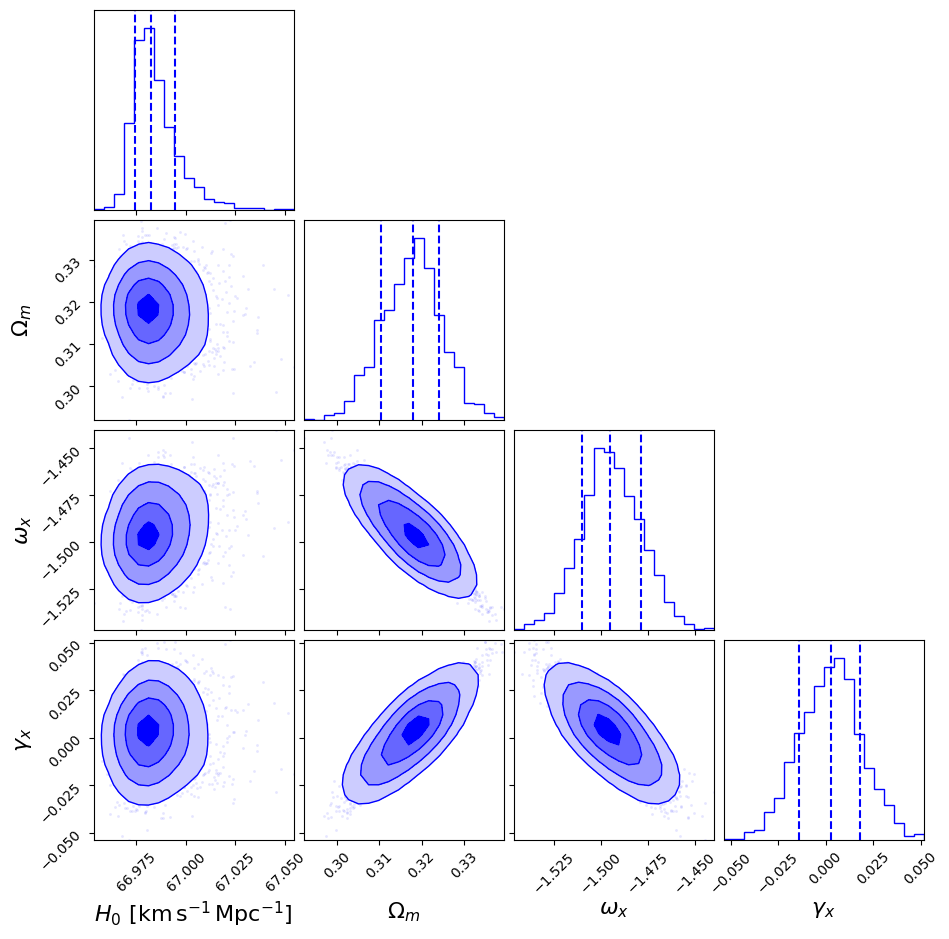}
    \end{minipage}
    
    \caption{The same as \autoref{f1}, but for the $\gamma_x$IDE model.}
    \label{f2}
\end{figure}

The free parameters in this model are $H_0$, $\Omega_m$, $\omega_x$, and $\gamma_x$. Using the MCMC method, we obtained the best-fit values and posterior distributions of these parameters by fitting the model to different FRB datasets.
The results are presented in \autoref{f2}, which present the posterior distributions derived from three datasets: the real FRB sample, a mock sample of 2,500 simulated FRBs, and a mock sample of 10,000 simulated FRBs, providing constraints on the $\gamma_x$ IDE model.

Using 86 localized FRBs, we obtain the following best-fit cosmological parameters: $H_0 = 81.03^{+5.24}_{-4.88} \, \mathrm{km\,s^{-1}\,Mpc^{-1}}$ and $\gamma_x = -0.80^{+0.75}_{-0.80}$ (68\% confidence level). but that $\omega_x$ is poorly constrained. Because the prior on $\omega_x$ is uniform over the interval $[-2, 0]$, only a median value of $\thicksim -1.03$ can be estimated. The best-fit value of $\gamma_x$ indicates that the energy is transferred from dark matter to dark energy, and the coincidence problem is not alleviated.

As in previous cases, we record the best-fit log-likelihood $\ln \mathcal{L}_{\rm max}$ and the number of free parameters $k = 4$ for BIC evaluation in Section~\ref{sec:Analysis}.

While current constraints based on 86 localized FRBs yield a Hubble constant of $H_0 = 81.03^{+5.24}_{-4.88}~\mathrm{km\,s^{-1}\,Mpc^{-1}}$ with a relative error of 6.2\%, the intermediate simulated sample significantly reduces this uncertainty to $\pm 0.25~\mathrm{km\,s^{-1}\,Mpc^{-1}}$. The large sample further compresses this uncertainty to $\pm 0.01~\mathrm{km\,s^{-1}\,Mpc^{-1}}$, underscoring the critical role of increased data volume in suppressing statistical fluctuations.

Similarly, the dark energy equation of state parameter $\omega_x$, initially constrained to $-1.03^{+0.65}_{-0.63}$ with an absolute uncertainty of $\pm 0.64$, sees its uncertainty reduced to $\pm 0.07$ for the intermediate sample and to $\pm 0.016$ for the large sample, demonstrating the potential of FRBs for precise cosmological parameter estimation.

For the coupling parameter $\gamma_x$, the broad observational constraint of $-0.80^{+0.75}_{-0.80}$ ($\pm 0.78$) narrows significantly to $\pm 0.018$ with the intermediate sample. This tightens further to $\pm 0.016$ for the large sample, indicating improved parameter stability but also highlighting the limited gains achievable through increased sample size alone due to parameter degeneracies. The slight posterior deviation of $\omega_x$ from its standard cosmological value likely results from residual parameter degeneracies that can shift best-fit estimates despite reduced statistical uncertainties.

\subsection{The $\xi$IDE Model}
In the $\xi$IDE model, the interaction term between dark matter and dark energy is expressed as $Q_3 = \frac{-(1-\Omega_m)(\xi + 3\omega_x)}{1 - \Omega_m + \Omega_m(1+z)^\xi} H\rho_m.$ The corresponding dimensionless Hubble parameter is then given by

\begin{equation}\label{16}
    E^2(z) = (1+z)^3 \left[ \Omega_m + (1 - \Omega_m)(1+z)^{-\xi} \right]^{-3\omega_x/\xi} \tag{16}
\end{equation}

\par
\begin{figure}[H]
    \centering
    \begin{minipage}[t]{0.45\textwidth}
        \centering
        \includegraphics[width=\linewidth]{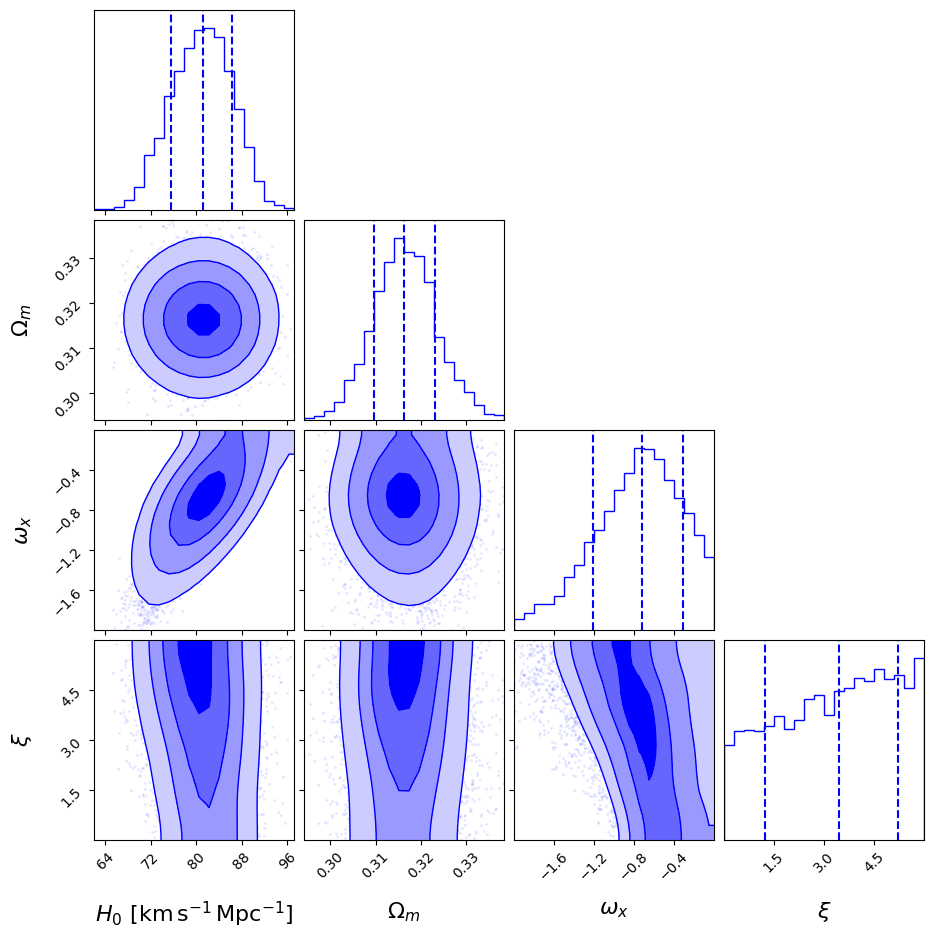}
    \end{minipage}
    
    \vspace{0.8em}  
    
    \begin{minipage}[t]{0.45\textwidth}
        \centering
        \includegraphics[width=\linewidth]{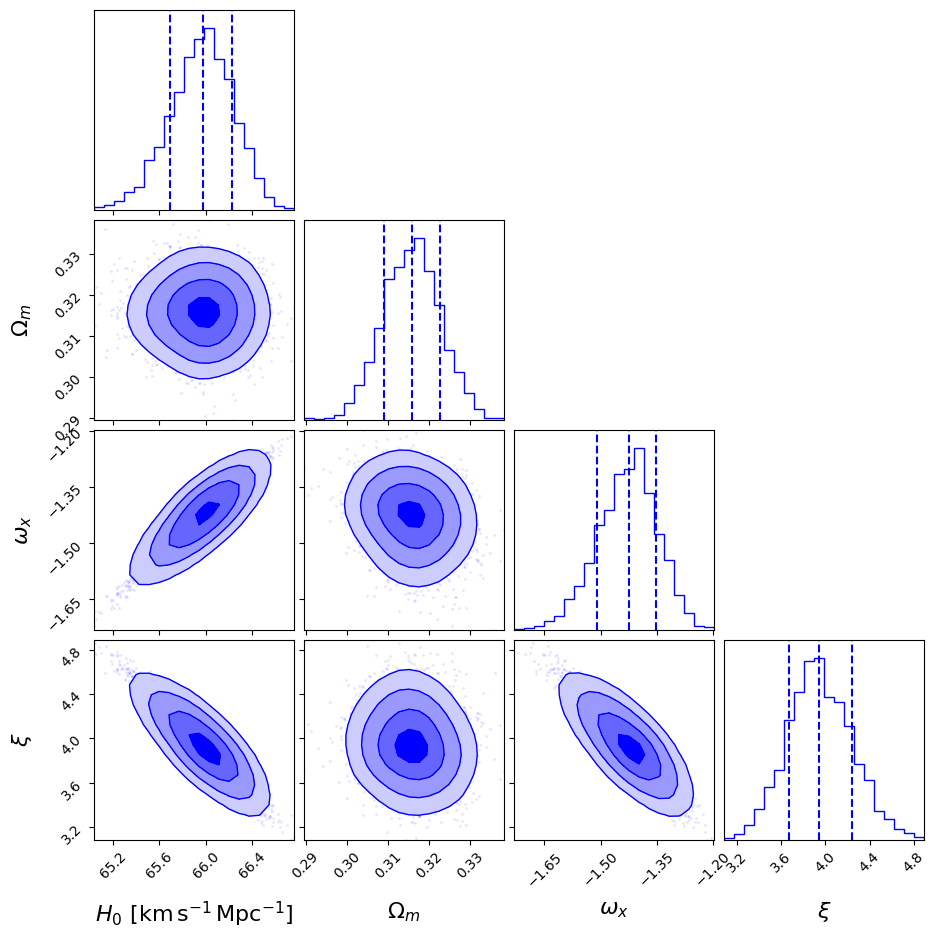}
    \end{minipage}
    \hfill
    \begin{minipage}[t]{0.45\textwidth}
        \centering
        \includegraphics[width=\linewidth]{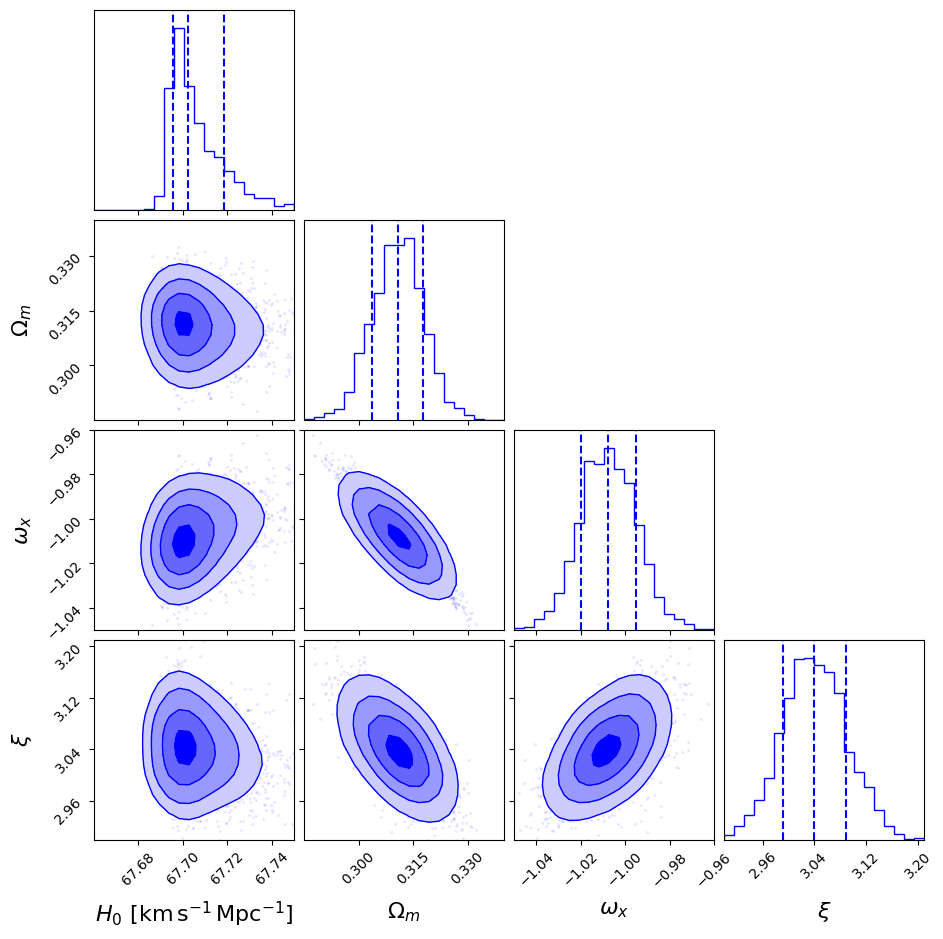}
    \end{minipage}
    
    \caption{The same as \autoref{f1}, but for the $\xi$IDE model.}
    \label{f3}
\end{figure}

 The free parameters are $H_0$, $\Omega_m$, $\omega_x$, and $\xi$. We employed the MCMC method to derive the best-fit values and posterior distributions of these parameters by fitting the model to various FRB datasets. The results are shown in \autoref{f3}.

For the real FRB sample, the best-fit values are 
\(H_0 = 81.14^{+5.11}_{-5.60}\ \mathrm{km\ s^{-1}\ Mpc^{-1}}\) and \(\omega_x = -0.72^{+0.41}_{-0.48}\) (68.3\% confidence level). 
The parameter \(\omega_x\) is poorly constrained, with the posterior distribution limited by the uniform prior over the interval \([-2, 0]\), allowing only a rough estimate of the median value at \(\omega_x \sim 3.44\). A strong degeneracy exists between \(\omega_x\) and \(\xi\), as both contribute to the effective coupling parameter 
\(\gamma = -(\xi + 3\omega_x)\), which governs the energy transfer direction. 
A rough estimate of the central value yields \(\gamma \sim -1.27\), suggesting a mild energy transfer from dark matter to dark energy, and indicating that the coincidence problem is not alleviated.

As in previous sections, the best-fit log-likelihood \(\ln \mathcal{L}_{\rm max}\) and the number of free parameters \(k=4\) are reported for model selection via the IC detailed in Section~\ref{sec:Analysis}.

Using simulated datasets, the constraints tighten significantly as the sample size increases. For a mock sample of 2,500 FRBs, the parameter uncertainties shrink to \( H_0 = 65.98^{+0.25}_{-0.29}\ \mathrm{km\,s^{-1}\,Mpc^{-1}} \), \( \omega_x = -1.42^{+0.07}_{-0.09} \), and \( \xi = 3.94^{+0.30}_{-0.27} \). Expanding the sample to 10,000 FRBs further improves the precision, yielding \( H_0 = 67.70^{+0.01}_{-0.01}\ \mathrm{km\,s^{-1}\,Mpc^{-1}} \), \( \omega_x = -1.008^{+0.01}_{-0.01} \), and \( \xi = 3.04^{+0.05}_{-0.04} \). These results demonstrate a clear convergence toward the \(\Lambda\)CDM values, particularly \(\xi = 3\).

\section{Analysis}
\label{sec:Analysis}
To systematically evaluate the performance of the three IDE models in fitting cosmological parameters, we employ IC for model comparison. These criteria balance model fit quality and complexity to identify the most statistically robust model. Specifically, we use three criteria: the Bayesian Information Criterion (BIC)\upcite{z2}, the Akaike Information Criterion (AIC)\upcite{z1}, and the Kullback Information Criterion (KIC)\upcite{z3}. Previous studies have extensively investigated the application of BIC, AIC, and KIC in cosmological contexts\upcite{z4,z5,z6,lin2024time}.
\par
The BIC is given by
\begin{equation}\label{17}
    \mathrm{BIC} = \mathrm{-2\ln }\mathcal{L}_{\mathrm{max}}+k\mathrm{\ln }N \tag{17}
\end{equation}
the AIC is defined as
\begin{equation}\label{18}
    \mathrm{AIC} = \mathrm{-2\ln }\mathcal{L}_{\mathrm{max}}+2k \tag{18}
\end{equation}
and the KIC is defined as
\begin{equation}\label{19}
    \mathrm{KIC} = \mathrm{-2\ln }\mathcal{L}_{\mathrm{max}}+3k \tag{19}
\end{equation}
Here, \(\mathcal{L}_{\max}\) denotes the maximum likelihood, \(k\) represents the number of parameters, and \(N\) denotes the number of data points. For Gaussian errors, \(\chi^2_{\min} = -2 \ln \mathcal{L}_{\max}\). We compute \(\chi^2_{\min}\) and calculate the corresponding AIC, BIC, and KIC values, which are presented in \autoref{2t}.

\par In the pairwise model comparison, the likelihood of model \( M_{\alpha} \) with information criterion value \( IC_{\alpha} \) is given by\upcite{wei2014comparison}.

\begin{equation}
P(M_{\alpha}) = \frac{\exp\left(-\frac{\mathrm{IC_{\alpha}}}{2}\right)}{\exp\left(-\frac{\mathrm{IC_1}}{2}\right) + \exp\left(-\frac{\mathrm{IC_2}}{2}\right)} \tag{20}
\end{equation}

The difference \(\Delta \mathrm{IC} = \mathrm{IC}_2 - \mathrm{IC}_1\) quantifies the degree to which model \(M_1\) is favored over model \(M_2\) \upcite{schwarz1978estimating,akaike2003new,cavanaugh2004criteria}. As an illustrative example, consider the comparison between the $\gamma_m$IDE and $\gamma_x$IDE models. Based on all three information criteria and using the real FRB sample, the probability that $\gamma_m$IDE is the preferred model is $P(M_1) \approx 50.4\%$, compared to $P(M_2) \approx 49.6\%$ for $\gamma_x$IDE. This result indicates that $\gamma_m$IDE yields slightly lower IC values than $\gamma_x$IDE, though the difference is not statistically significant. A similar marginal difference is also observed between the $\gamma_x$IDE and $\xi$IDE models.

\begin{figure}[H]
    \centering
    \begin{minipage}[c]{0.4\textwidth}
        \centering
        \includegraphics[width=\linewidth]{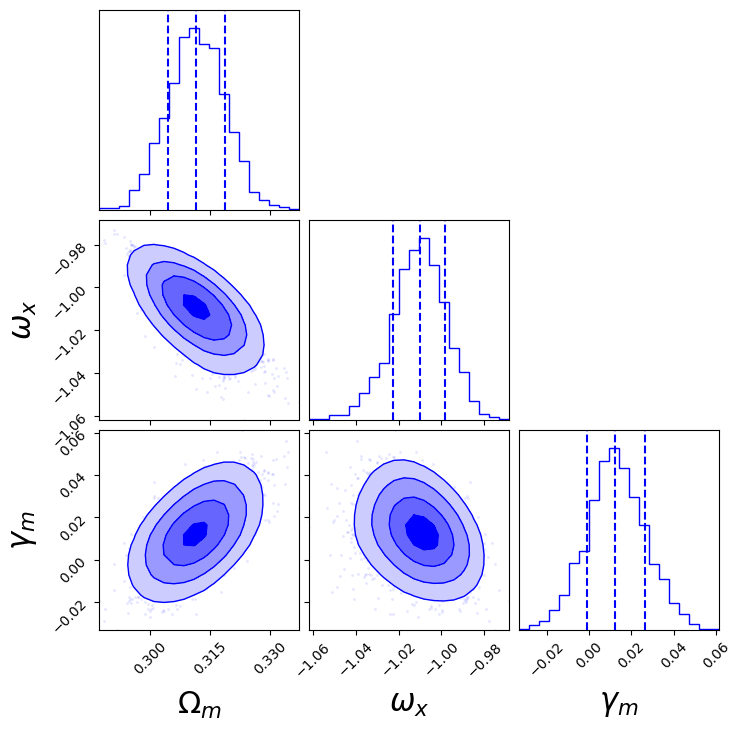}
    \end{minipage}
    \hspace{0.05\textwidth}
    \begin{minipage}[c]{0.4\textwidth}
        \centering
        \includegraphics[width=\linewidth]{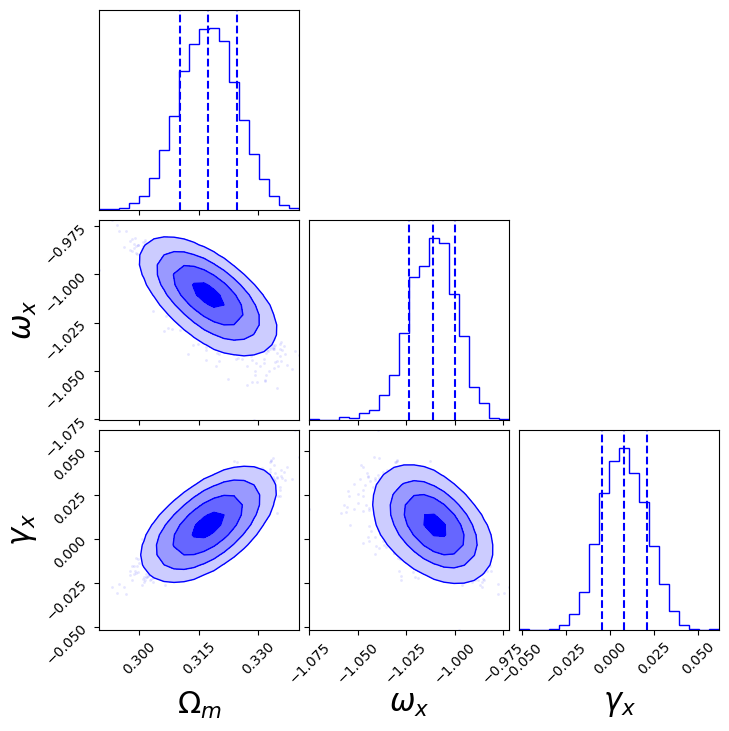}
    \end{minipage}
    \caption{Constraints on the remaining parameters of the $\gamma_m$ and $\gamma_x$ models using 2500 simulated FRBs, with $H_0$ fixed at 67.4 $\mathrm{km\ s^{-1}\ Mpc^{-1}}$}. The left panel shows the results for the $\gamma_m$ model, while the right panel corresponds to the $\gamma_x$ model.
    \label{f4}
\end{figure}

\begin{table}[H]
    \centering
    \caption{IC values for the three models based on real data}
    \label{2t}
    \begin{tabular}{lccc}
        \toprule[1.0pt]
        \setlength{\arrayrulewidth}{1.0pt}
        IC & $\gamma_m$IDE model & $\gamma_x$IDE model& $\xi$IDE model\\
        \hline

             $\chi^2_{\mathrm{min}}$ & 1136.319 & 1136.347 &1136.317\\
             BIC                      & 1154.137 & 1154.165 & 1154.134\\
             AIC                      & 1144.319 & 1144.347 &1144.317\\
             KIC                      & 1148.319 & 1148.347 &1148.317\\

        \bottomrule[1.0pt]
    \end{tabular}
\end{table}
Finally, for the \(\gamma_m\)IDE and \(\gamma_x\)IDE models, the best-fit values of \(\omega_x\) tend to be smaller than the fiducial value, likely due to parameter correlations. To verify this, we fix \(H_0 = 67.4\ \mathrm{km\ s^{-1}\ Mpc^{-1}}\) and constrain the remaining three parameters (\(\Omega_m\), \(\gamma_m\), and \(\omega_x\)) with 2,500 simulated samples, following the same procedure as the previous four-parameter model. As shown in \autoref{f4}, the best-fit values for the \(\gamma_m\)IDE model are \(\omega_x = -1.01^{+0.012}_{-0.013}\); for the \(\gamma_x\)IDE model, they are \(\omega_x = -1.011^{+0.012}_{-0.012}\). All parameters recover their fiducial values within the \(1\sigma\) uncertainties.

\section{Conclusion}\label{s6}
In this paper, we explore the potential of FRBs to constrain cosmological parameters and investigate IDE models. We leverage the unique characteristics of FRBs, particularly their DMs, and redshifts, to perform a comprehensive analysis utilizing both observational and simulated FRB datasets.

\par
We focus on three IDE models: the \(\gamma_m\)IDE, \(\gamma_x\)IDE, and \(\xi\)IDE models. Using Bayesian MCMC methods, we systematically evaluate the constraints on key cosmological parameters, namely the Hubble constant \(H_0\), the matter density parameter \(\Omega_m\), the dark energy equation of state parameter \(\omega_x\), and the interaction parameters \(\gamma_m\), \(\gamma_x\), and \(\xi\).

First, we investigate three IDE models—\(\gamma_m\)IDE, \(\gamma_x\)IDE, and \(\xi\)IDE—using current localized FRB data alongside extended mock FRB samples. Based on observed FRB data, we find that the interaction term between dark energy and dark matter is consistent with zero within uncertainties, in agreement with previous studies constraining interaction parameters. In the \(\gamma_m\)IDE model, the interaction parameter \(\gamma_m = 0\) falls within the \(1\sigma\) confidence interval, consistent with the standard \(\Lambda\)CDM model. However, the parameters \(\omega_x\) in the \(\gamma_x\)IDE model and \(\xi\) in the \(\xi\)IDE model are weakly constrained, resulting in posterior distributions that are nearly uniform across their prior ranges. This suggests that the current FRB dataset lacks sufficient constraining ability to confirm or exclude dark energy-dark matter interactions characterized by coupling terms proportional to matter density, dark energy density, or their ratio parameterized as a power-law of the scale factor.

With ongoing and upcoming radio telescopes, the number of localized FRBs is expected to increase significantly soon, and the redshift coverage will likely extend up to \( z \approx 3 \). To anticipate the constraining power of future data, we perform the same analysis using two mock datasets comprising 2,500 and 10,000 FRBs. We find that constraints on \(H_0\), \(\omega_x\), and \(\xi\) significantly improve with increasing sample size. However, the benefits for \(\gamma_m\) and \(\gamma_x\) exhibit diminishing returns, indicating a saturation in statistical gain. In the \(\xi\)IDE model, the parameters accurately recover their fiducial values \((H_0 = 67.4\ \mathrm{km\ s^{-1}\ Mpc^{-1}}, \omega_x = -1, \xi = 3)\) with increasing sample size, while in the \(\gamma_m\)IDE and \(\gamma_x\)IDE models, the parameter \(\omega_x\) continues to deviate from its fiducial value, likely owing to residual parameter degeneracies.

Moreover, we provide three information criteria (AIC, BIC, and KIC) for each model in \autoref{2t}. The current observational data are insufficient to definitively discriminate among the three IDE models. Under the present FRB observations, the $\xi$IDE model yields slightly lower IC values than the other two models, but the differences are not statistically significant.

Lastly, we investigate the impact of fixing \(H_0 = 67.4\ \mathrm{km\ s^{-1}\ Mpc^{-1}}\) on parameter recovery using the \(\gamma_m\)IDE and \(\gamma_x\)IDE models with 2,500 mock FRBs. The resulting posterior distributions demonstrate that the parameters \(\omega_x\) and \(\gamma\) are more accurately recovered and consistent with the fiducial values, confirming the presence of residual degeneracies that can be mitigated by fixing certain parameters.

In conclusion, our results highlight the potential of FRBs in testing interacting dark energy scenarios. While current FRB data alone are insufficient to constrain dark energy–dark matter interactions, simulations suggest that future large and well-localized FRB samples could serve as robust cosmological probes, improving parameter estimation and enabling discrimination between interacting and non-interacting dark energy models.

\section{Acknowledgments}
This work was supported by the National Natural Science Foundation of China (Grants No. 12105032 and 12203010), the Science and Technology Research Program of Chongqing Municipal Education Commission (Grant No. KJQN202200633), and the Natural Science Foundation of Chongqing (Grants No. cstc2021jcyj-msxmX0481, cstc2021jcyj-msxmX0553, and CSTB2023NSCQ-MSX1048).
\begin{center}
    \Large\bfseries APPENDIX
\end{center}
\begin{longtable}{lcccccccc}
    \caption{Properties of 86 localized FRBs} \label{tab1} \\
    \toprule[1.0pt]
    Names & Redshift & $\mathrm{DM_{obs}}$  & $\rm R.A.$ & Decl.& $\mathrm{DM_{ISM}^{MW}}$  & Host Type& Reference$\null\qquad\;$\\
         &          & $(\mathrm{pc\;cm^{-3}})$  & $(\mathrm{deg,J2000})$ & $(\mathrm{deg,J2000})$   &$(\mathrm{pc\;cm^{-3}})$     &  &  \\
    \hline
    \endfirsthead

    \multicolumn{9}{c}%
    {{\bfseries Table \thetable\ continued from previous page}} \\
    \toprule[1.0pt]
    Names & Redshift & $\mathrm{DM_{obs}}$  & $\rm R.A.$ & Decl.& $\mathrm{DM_{ISM}^{MW}}$  & Host Type& Reference$\null\qquad\;$\\
         &          & $(\mathrm{pc\;cm^{-3}})$  & $(\mathrm{deg,J2000})$ &  $(\mathrm{deg,J2000})$  &  $(\mathrm{pc\;cm^{-3}})$&   &  &  \\
    \hline
    \endhead

    \bottomrule[1pt]
    \multicolumn{9}{r}{{Continued on next page}} \\
    \endfoot

    \hline
    \endlastfoot
        FRB121102A&0.19273&557&82.9946&33.1479&188.4&1& \cite{2017Natur.541...58C,2017ApJ...834L...7T}\\
        FRB180301A&0.3304&552&93.2268&4.6711&151.7&1& \cite{2022AJ....163...69B}\\
        FRB180814A&0.068&190.9&65.6833&73.6644&87.6&2&\cite{2023ApJ...950..134M}\\
        FRB180916B&0.0337&349.349&29.5031&65.7168&199&2&\cite{2020Natur.577..190M}\\
        FRB180924B&0.3212&361.42&326.1053&-40.9&40.5&3&\cite{2023ApJ...954...80G,2019Sci...365..565B,2020Natur.581..391M,2020ApJ...895L..37B,2024arXiv240802083S}\\
        FRB181112A&0.4755&589.27&327.3485&-52.9709&41.7&3&\cite{2019Sci...366..231P,2020Natur.581..391M,2020ApJ...895L..37B,2024arXiv240802083S}\\
        FRB181220A&0.02746&208.66&348.6982&48.3421&118.5&3&\cite{2024ApJ...971L..51B}\\
        FRB181223C&0.03024&111.61&180.9207&27.5476&19.9&3&\cite{2024ApJ...971L..51B}\\
        FRB190102C&0.2912&364.5&322.4157&-79.4757&57.4&3&\cite{2020Natur.581..391M,2020ApJ...895L..37B,2024arXiv240802083S}\\
        FRB190110C&0.12244&221.6&249.3185&41.4434&37.1&2&\cite{2024ApJ...961...99I}\\
        FRB190303A&0.064&223.2&207.9958&48.1211&29.8&2&\cite{2023ApJ...950..134M}\\
        FRB190418A&0.07132&182.78&65.8123&16.0738&70.2&3&\cite{2024ApJ...971L..51B}\\
        FRB190520B&0.2418&1204.7&240.5178&-11.2881&60.2&1&\cite{2023ApJ...954...80G,2022Natur.606..873N}\\
        FRB190523A&0.66&760.8&207.065&72.4697&37.2&3&\cite{2019Natur.572..352R}\\
        FRB190608B&0.11778&338.7&334.0199&-7.8983&37.3&3&\cite{2023ApJ...954...80G,2020Natur.581..391M,2020ApJ...895L..37B,2024arXiv240802083S}\\
        FRB190611B&0.3778&321.4&320.7456&-79.3976&57.8&3&\cite{2020ApJ...903..152H,2020Natur.581..391M,2024arXiv240802083S}\\
        FRB190614D&0.6&959.2&65.0755&73.7067&87.8&3&\cite{2020ApJ...899..161L}\\
        FRB190711A&0.522&593.1&329.4193&-80.358&56.5&1&\cite{2020ApJ...903..152H,2020Natur.581..391M,2024arXiv240802083S}\\
        FRB190714A&0.2365&504.13&183.9797&-13.021&38.5&3&\cite{2020ApJ...903..152H,2024arXiv240802083S}\\
        FRB191001A&0.234&506.92&323.3513&-54.7478&44.2&3&\cite{2020ApJ...903..152H,2024arXiv240802083S}\\
        FRB191106C&0.10775&332.2&199.5801&42.9997&25&2&\cite{2024ApJ...961...99I}\\
        FRB191228A&0.2432&297.5&344.4304&-28.5941&32.9&3&\cite{2022AJ....163...69B,2024arXiv240802083S}\\
        FRB200223B&0.06024&201.8&8.2695&28.8313&45.6&2&\cite{2024ApJ...961...99I}\\
        FRB200430A&0.1608&380.1&229.7064&12.3763&27.2&3&\cite{2020ApJ...903..152H,2024arXiv240802083S}\\
        FRB200906A&0.3688&577.8&53.4962&-14.0832&35.8&3&\cite{2022AJ....163...69B,2024arXiv240802083S}\\
        FRB201123A&0.0507&433.55&263.67&-50.76&251.7&1&\cite{2022MNRAS.514.1961R}\\
        FRB201124A&0.098&413.52&77.0146&26.0607&139.9&2&\cite{2021ApJ...919L..23F}\\
        FRB210117A&0.2145&729.1&339.9792&-16.1515&34.4&3&\cite{2023ApJ...954...80G,2024arXiv240802083S}\\
        FRB210320C&0.2797&384.8&204.4608&-16.1227&39.3&3&\cite{2023ApJ...954...80G,2024arXiv240802083S}\\
        FRB210410D&0.1415&571.2&326.0863&-79.3182&56.2&3&\cite{2023ApJ...954...80G,2023MNRAS.524.2064C}\\
        FRB210603A&0.1772&500.147&10.2741&21.2263&39.5&3&\cite{2023arXiv230709502C}\\
        FRB210807D&0.1293&251.9&299.2214&-0.7624&121.2&3&\cite{2023ApJ...954...80G,2024arXiv240802083S}\\
        FRB211127I&0.0469&234.83&199.8082&-18.8378&42.5&3&\cite{2023ApJ...954...80G,2023ApJ...949...25G,2024arXiv240802083S}\\
        FRB211203C&0.3439&636.2&204.5625&-31.3801&63.7&3&\cite{2023ApJ...954...80G,2024arXiv240802083S}\\
        FRB211212A&0.0707&206&157.3509&1.3609&38.8&3&\cite{2023ApJ...954...80G,2024arXiv240802083S}\\
        FRB220105A&0.2785&583&208.8039&22.4665&22&3&\cite{2023ApJ...954...80G,2024arXiv240802083S}\\
        FRB220204A&0.4012&612.584&274.2263&69.7225&50.7&3&\cite{2024ApJ...964..131S,2024Natur.635...61S,2024arXiv240916952C}\\
        FRB220207C&0.04304&262.38&310.1995&72.8823&76.1&3&\cite{2024ApJ...964..131S,2024arXiv240916952C}\\
        FRB220307B&0.248123&499.27&350.8745&72.1924&128.2&3&\cite{2024ApJ...964..131S,2024arXiv240916952C}\\
        FRB220310F&0.477958&462.24&134.7204&73.4908&46.3&3&\cite{2024ApJ...964..131S,2024arXiv240916952C}\\
        FRB220418A&0.622&623.25&219.1056&70.0959&36.7&3&\cite{2024ApJ...964..131S,2024arXiv240916952C}\\
        FRB220501C&0.381&449.5&352.3792&-32.4907&30.6&3&\cite{2024arXiv240802083S,2024Natur.635...61S}\\
        FRB220506D&0.30039&396.97&318.0448&72.8273&84.6&3&\cite{2024ApJ...964..131S,2024Natur.635...61S,2024arXiv240916952C}\\
        FRB220509G&0.0894&269.53&282.67&70.2438&55.6&3&\cite{2024ApJ...971L..51B,2024ApJ...964..131S,2024arXiv240916952C}\\
        FRB220529A&0.1839&246&19.1042&20.6325&40&1&\cite{2024arXiv241003994G}\\
        FRB220610A&1.016&1458.15&351.0732&-33.5137&31&3&\cite{2024arXiv240802083S}\\
        FRB220717A&0.36295&637.34&293.3042&-19.2877&118.3&3&\cite{2024MNRAS.532.3881R}\\
        FRB220725A&0.1926&290.4&353.3152&-35.9902&30.7&3&\cite{2024arXiv240802083S}\\
        FRB220726A&0.3619&686.232&73.94567&69.9291&89.5&3&\cite{2024ApJ...964..131S,2024Natur.635...61S,2024arXiv240916952C}\\
        FRB220825A&0.241397&651.24&311.9815&72.585&78.5&3&\cite{2024ApJ...964..131S,2024arXiv240916952C}\\
        FRB220831A&0.262&1146.25&338.6955&70.5384&126.8&3&\cite{2024arXiv240916952C}\\
        FRB220912A&0.0771&219.46&347.2704&48.7071&125.2&2&\cite{2023ApJ...949L...3R}\\
        FRB220914A&0.1139&631.28&282.0568&73.3369&54.7&3&\cite{2024ApJ...964..131S,2024arXiv240916952C}\\
        FRB220918A&0.491&656.8&17.5921&70.8113&153.1&3&\cite{2024arXiv240802083S}\\
        FRB220920A&0.158239&314.99&240.2571&70.9188&39.9&3&\cite{2024ApJ...964..131S,2024arXiv240916952C}\\
        FRB221012A&0.284669&441.08&280.7987&70.5242&54.3&3&\cite{2024ApJ...964..131S,2024arXiv240916952C}\\
        FRB221029A&0.975&1391.75&141.9634&72.4523&43.8&3&\cite{2024ApJ...964..131S,2024Natur.635...61S,2024arXiv240916952C}\\
        FRB221101B&0.2395&491.554&342.2162&70.6812&131.2&3&\cite{2024ApJ...964..131S,2024Natur.635...61S,2024arXiv240916952C}\\
        FRB221106A&0.2044&343.8&56.7048&-25.5698&34.8&3&\cite{2024ApJ...964..131S,2024arXiv240802083S}\\
        FRB221113A&0.2505&411.027&71.411&70.3074&91.7&3&\cite{2024ApJ...964..131S,2024Natur.635...61S,2024arXiv240916952C}\\
        FRB221116A&0.2764&643.448&21.2102&72.6539&132.3&3&\cite{2024Natur.635...61S,2024arXiv240916952C}\\
        FRB221219A&0.553&706.708&257.6298&71.6268&44.4&3&\cite{2024ApJ...964..131S,2024Natur.635...61S,2024arXiv240916952C}\\
        FRB230124A&0.0939&590.574&231.9163&70.9681&38.6&3&\cite{2024ApJ...964..131S,2024Natur.635...61S,2024arXiv240916952C}\\
        FRB230307A&0.2706&608.854&177.7813&71.6956&37.6&3&\cite{2024ApJ...964..131S,2024Natur.635...61S,2024arXiv240916952C}\\
        FRB230501A&0.3015&532.471&340.0272&70.9222&125.7&3&\cite{2024ApJ...964..131S,2024arXiv240916952C}\\
        FRB230521B&1.354&1342.9&351.036&71.138&138.8&3&\cite{2024arXiv240802083S,2024arXiv240916952C}\\
        FRB230526A&0.157&361.4&22.2326&-52.7173&31.9&3&\cite{2024arXiv240802083S}\\
        FRB230626A&0.327&452.723&235.6296&71.1335&39.3&3&\cite{2024ApJ...964..131S,2024Natur.635...61S,2024arXiv240916952C}\\
        FRB230628A&0.127&344.952&166.7867&72.2818&39&3&\cite{2024ApJ...964..131S,2024Natur.635...61S,2024arXiv240916952C}\\
        FRB230708A&0.105&411.51&303.1155&-55.3563&60.3&3&\cite{2024arXiv240802083S}\\
        FRB230712A&0.4525&587.567&167.3585&72.5578&39.2&3&\cite{2024ApJ...964..131S,2024Natur.635...61S,2024arXiv240916952C}\\
        FRB230814A&0.553&696.4&335.9748&73.0259&104.8&3&\cite{2024arXiv240916952C}\\
        FRB230902A&0.3619&440.1&52.1398&-47.3335&34.1&3&\cite{2024arXiv240802083S}\\
        FRB231120A&0.0368&437.737&143.984&73.2847&43.8&3&\cite{2024ApJ...964..131S,2024Natur.635...61S,2024arXiv240916952C}\\
        FRB231123B&0.2621&396.857&242.5382&70.7851&40.3&3&\cite{2024ApJ...964..131S,2024Natur.635...61S,2024arXiv240916952C}\\
        FRB231220A&0.3355&491.2&123.9087&73.6599&49.9&3&\cite{2024arXiv240916952C}\\
        FRB231226A&0.1569&329.9&155.3638&6.1103&38.1&3&\cite{2024arXiv240802083S}\\
        FRB240114A&0.13&527.65&321.9161&4.3292&49.7&1&\cite{2024MNRAS.533.3174T}\\
        FRB240119A&0.376&483.1&224.4672&71.6118&38&3&\cite{2024arXiv240916952C}\\
        FRB240123A&0.968&1462&68.2625&71.9453&90.2&3&\cite{2024arXiv240916952C}\\
        FRB240201A&0.042729&374.5&149.9056&14.088&38.6&3&\cite{2024arXiv240802083S}\\
        FRB240210A&0.023686&283.73&8.7796&-28.2708&28.7&3&\cite{2024arXiv240802083S}\\
        FRB240213A&0.1185&357.4&166.1683&74.0754&40&3&\cite{2024arXiv240916952C}\\
        FRB240215A&0.21&549.5&268.4413&70.2324&47.9&3&\cite{2024arXiv240916952C}\\
        FRB240229A&0.287&491.15&169.9835&70.6762&38&3&\cite{2024arXiv240916952C}\\
        FRB240310A&0.127&601.8&17.6219&-44.4394&30.1&3&\cite{2024arXiv240802083S}\\
        \bottomrule[1.0pt]
    
\end{longtable}

\bibliography{References}

\end{document}